# INFLUENCE OF HIGH PRESSURE ON THE REMARKABLE ITINERANT ELECTRON BEHAVIOUR IN $Y_{0.7}Er_{0.3}Fe_2D_{4.2}$ COMPOUND

Z. Arnold[1], O. Isnard[2], V. Paul-Boncour[3*]

[1]*Institute of Physics AS CR, v.v.i., Na Slovance 2, 18221 Prague 8, Czech Republic*
[2]*Université Grenoble Alpes, Institut Néel, CNRS, BP166X, 38042 Grenoble Cédex 9, France*
[3] *Université Paris-Est Créteil, CNRS, ICMPE, UMR7182, F-94320 Thiais, France*

ORCID
Z. Arnold–0000-0002-6271-9569
O. Isnard-0000-0002-7930-4378
V. Paul-Boncour–0000-0002-0601-7802

*Corresponding author, email: valerie.paul-boncour@cnrs.fr*

**Abstract**

Monoclinic $Y_{0.7}Er_{0.3}Fe_2D_{4.2}$ compound exhibits unusual magnetic properties with different field induced magnetic transitions. The deuteride is ferrimagnetic at low temperature and the Er and Fe sublattices present magnetic transitions at different temperatures. The Er moments are ordered below $T_{Er}$=55 K, whereas the Fe moments remain ferromagnetically coupled up to $T_{M0}$ = 66 K. At $T_{M0}$ the Fe moments display a sharp ferromagnetic-antiferromagnetic transition (FM-AFM) through an itinerant electron metamagnetic (IEM) behaviour very sensitive to any volume change. $Y_{0.7}Er_{0.3}Fe_2D_{4.2}$ becomes paramagnetic above $T_N$=125 K. The pressure dependence of $T_{Er}$ and $T_{M0}$ have been extracted from magnetic measurements under hydrostatic pressure up to 0.49 GPa. Both temperatures decrease linearly upon applied pressure with $dT_{Er}/dP$=-126 and $dT_{M0}/dP$=-140 K.GPa$^{-1}$ for a field of $B$=0.03 T. Both magnetic Er and ferromagnetic Fe order disappear at $P$=0.44(4) GPa. However, under a larger applied field $B$=5 T, $dT_{M0}/dP$=-156 K.GPa$^{-1}$ whereas $dT_{Er}/dP$=-134 K.GPa$^{-1}$ showing a weaker sensitivity to pressure and magnetic field. At 2 K the decrease of the saturation magnetization under pressure can be attributed to a reduction of the mean Er moment due to canting and/or crystal field effect. Above $T_{M0}$ the magnetization curves display a metamagnetic behaviour from AFM to FM state, which is also very sensitive to the applied pressure. The transition field $B_{trans}$, which increases linearly upon heating, is shifted to lower temperature upon applied pressure with $\Delta T$=-17 K between 0 and 0.11 GPa. These results show a strong decoupling of the Er and Fe magnetic sublattices versus temperature, applied field and pressure.

## I. INTRODUCTION

$R$Fe$_2$ (R =Rare-Earth) Laves phase compounds have been widely studied for their magnetic properties[1], especially for their giant magnetostrictive properties[2-5], leading to various applications such as transducers, actuators or motors[6]. From the mechanical point of view, $R$Fe$_2$ bulk ingots can be decrepitated upon hydrogen insertion as used to prepare sintered Terfenol-D (Tb$_{0.3}$Dy$_{0.7}$Fe$_{1.9}$ composition) with optimized performances for applications[7,8]. The influence of hydrogen insertion on the magnetic and magnetostrictive properties of ($R$,Tb)Fe$_2$ compounds ($R$ = Er, Dy, Ho) has been also investigated showing a large cell volume increase, in some cases a rhombohedral distortion, a drastic lowering of $T_C$ and magnetization, and a large changes of magnetostrictive properties [9-13]. Hydrogen absorption was therefore investigated systematically in several $R$Fe$_2$ type compounds to understand more clearly the relationship between structural changes and magnetic properties. It was found that $R$Fe$_2$ compounds can absorb large hydrogen content, up to 5 H/f.u. [14,15] and that H insertion can induce many different and interesting structural and magnetic transitions [16-22].

Whereas the Fe magnetism is of simple ferrimagnetic type in cubic YFe$_2$ compound (considering an induced weak moment on Y site), the presence of hydrogen as interstitial element within the crystal lattice induces tremendous changes of both its crystal structure and physical properties [23-26]. Depending upon the H concentration several crystal structures have been observed at room temperature due to H order into interstitial sites [27,28]. In particular, a lowering of crystal symmetry is occurring from the cubic C15 Laves type down to a monoclinic structure for the YFe$_2$H$_{4.2}$ compound [29] and to an orthorhombic structure for YFe$_2$H$_5$ [15]. The magnetic properties are also found to be extremely sensitive to the H concentration since YFe$_2$H$_5$ is no longer magnetically ordered [30,31], whereas YFe$_2$H$_{4.2}$ present a complex magnetic behaviour due to the competition between ferromagnetic and antiferromagnetic exchange interactions [29,32]. YFe$_2$H$_{4.2}$ compound exhibits a ferromagnetic-antiferromagnetic (FM-AFM) transition at $T_{M0}$ =131 K, which has been proved to be very sensitive to the application of external parameters such as applied magnetic field, applied pressure [33], chemical substitution [34] or even (H,D) isotopic effect [29]. Above the FM-AFM transition a metamagnetic behaviour is observed with a linear increase of the transition field versus temperature [32]. An AFM-paramagnetic (PM) transition is observed at higher temperature above the Néel temperature $T_N$ =160 K.

The high sensitivity of the Fe magnetism to its local atomic environment in such compounds is demonstrated by the giant isotopic effect that has been reported recently as the FM-AFM transition temperature is shifted to $T_{M0}$ = 84 K (-47 K) in the YFe$_2$D$_{4.2}$ deuteride [32]. This difference of transition temperature has been related to cell volume reduction of 0.8 % of the deuteride compared to the hydride, due to the difference of zero-point amplitude of vibration of the hydrogen isotopes into the interstitial sites. The transition temperature can also be strongly reduced by an external pressure [33,35]. This FM-AFM transition is also associated to a large variation of the magnetic entropy showing interesting magnetocaloric effects. These YFe$_2$ related compounds constitute therefore an ideal playground to investigate the mechanism responsible for such unusual magnetic properties and offers the opportunity to go deeper in the understanding of the itinerant electron behaviour of Fe. Indeed, itinerant electron magnetism (IEM) has attracted much interest from both experimentalists and theoreticians over the last decades[36] including a revival due to the discovery of giant magnetocaloric effect in such Fe itinerant electron systems at the verge of the antiferromagnetic to ferromagnetic ordering [37].

Due to the high sensitivity of this IEM transition to the volume changes, the influence of the cell volume reduction not only under hydrostatic pressure but also by chemical substitution on the Y site by another rare-earth element of smaller radius have been investigated in Y$_{1-x}$Er$_x$Fe$_2$ hydrides and deuterides (0 < $x$ < 1) [34]. Two different types of field induced magnetic transitions have been observed for deuterides with $x$ = 0.3 and 0.5 by combining x-ray and neutron diffraction with magnetic measurements under high magnetic field [38,39]. In the present work we have decided to focus on the influence of the applied pressure on the Y$_{0.7}$Er$_{0.3}$Fe$_2$D$_{4.2}$ magnetic properties as the two magnetic transitions are occurring in separated temperature ranges and consequently can be better followed independently. This $x$ = 0.3 composition was also selected since both its crystal structure and magnetic properties were fully characterized by neutron diffraction and high magnetic field measurements.

The unusual magnetic properties of Y$_{0.7}$Er$_{0.3}$Fe$_2$D$_{4.2}$ [38] can be summarized as follows. It is ferrimagnetically ordered up to $T_{Er}$ = 55 K, temperature at which the Er sublattice magnetization vanishes. In the ground state, the Er and Fe magnetic moments are coupled antiparallel. A forced ferrimagnetic-ferromagnetic transition (Ferri-FM) is found up to $T_{Er}$ with a transition field $B_{trans}$ around 8 T. At $T_{M0}$ = 66 K, the Fe sublattice magnetization undergoes a first-order transition from ferromagnetic to antiferromagnetic state (FM-AFM) leading to an antiferromagnetic type ordering up to the Néel temperature $T_N$. It has been demonstrated previously that between $T_{M0}$ and $T_N$, Y$_{0.7}$Er$_{0.3}$Fe$_2$D$_{4.2}$ exhibits field induced magnetic transitions from AFM to FM state of the Fe sublattice. This AFM-FM transition of the Fe sublattice is typical of IEM behaviour and featured by a transition field $B_{trans}$, which increases linearly with the temperature. More details on the magnetic moment arrangement and its temperature and field dependence can be found in the magnetic phase diagram described in [38]. The AFM state is observed up to $T_N$ = 125 K, which is only 6 K smaller than that for $YFe_2D_{4.2}$ ($T_N$ = 131 K). On the contrary, the influence of Er for Y substitution on $T_{M0}$ is 3 times more pronounced with a reduction of 18 K compared to that of $YFe_2D_{4.2}$. Above $T_N$, $Y_{0.7}Er_{0.3}Fe_2D_{4.2}$ compound displays a weak spontaneous magnetization, but without long range magnetic order indicating rather a disordered magnetic state. Additionally, the crystal structure investigation revealed that the FM-AFM transition occurring at $T_{M0}$ in $Y_{0.7}Er_{0.3}Fe_2D_{4.2}$ compound is accompanied by a significant contraction of the unit cell volume [38].

Previous study of $Y_{1-x}Er_xFe_2$ hydrides and deuterides has shown that the variation of $T_{M0}$ versus cell volume displays significantly different slopes when the contraction is induced by applying a pressure on $YFe_2(H)D_{4.2}$ compounds or upon Y for Er substitution. This motivated us to combine the influence of applied pressure on the interesting magnetic properties of $Y_{0.7}Er_{0.3}Fe_2D_{4.2}$ compound. Aiming to study the effect of volume change on the character of magnetism and exchange interactions, we present below the influence of hydrostatic pressure up to 1 GPa on magnetic properties of polycrystalline $Y_{0.7}Er_{0.3}Fe_2D_{4.2}$ compound. The results will be analysed in the light of the knowledge of the crystal structure, as well as recent high magnetic field results published previously[34,38] as well as on other Laves phase compounds presenting an IEM behaviour. They will also be compared and discussed with the influence of applied pressure on the magnetic properties of $YFe_2(H,D)_{4.2}$ [33].

## II. EXPERIMENTAL

The synthesis of the $Y_{0.7}Er_{0.3}Fe_2$ intermetallic compound was performed by induction melting together the pure elements under purified argon atmosphere followed by thermal annealing of 4 weeks at 1073 K in evacuated quartz ampoule. The mean sample composition analysed by Electron Probe Micro Analysis (EPMA from CAMECA) is $Y_{0.68(2)}Er_{0.27(2)}Fe_2$.

The elaboration of the corresponding deuteride has been performed by solid-gas reaction with deuterium gas using the Sievert method. The synthesis procedures are described in more details in [38]. The deuterium content was found to be 4.15±0.05 D/f.u. as estimated by a volumetric method and confirmed by neutron diffraction. The deuteride was quenched into liquid nitrogen and slowly heated under air up to room temperature to passivate the surface and avoid further deuterium desorption. The sample quality has been checked at room temperature before and after the deuterium insertion by means of X-ray powder diffraction technique using the Cu $K_\alpha$ radiation. $Y_{0.7}Er_{0.3}Fe_2$ compound is found to be single phase crystallizing in cubic

C15 crystal structure ($Fd\bar{3}m$ space group) structure with $a$ = 7.334(1) Å. $Y_{0.7}Er_{0.3}Fe_2D_{4.2}$ is monoclinic ($Pc$ space group) with cell parameters and atomic positions refined by neutron diffraction reported in [38].

The magnetic properties were determined in the SQUID magnetometer (Quantum Design Co.) in temperature range 5 – 300 K with magnetic field up to 7 T. The magnetization curves recorded at high hydrostatic pressure were measured using a miniature CuBe pressure cell of piston-cylinder type in pressure ranging up to 1 GPa (10 kbar). A mixture of mineral oils is used as a pressure transmitting medium and a piece of lead as pressure reference. Indeed, the pressure inside the cell was determined at low temperatures using the known pressure dependence of the critical temperature of the superconducting state of the pure Pb (5N) sample [40].

The evolution of the magnetic transition temperatures under different pressures were determined from temperature dependence of the isofield magnetization curves. $T_{Er}$ was defined as the maximum of magnetization vs. temperature curves; and the FM-AFM transition temperature was defined as the inflection point of low field magnetization vs. temperature curve. The transition fields $B_{trans}$ are determined as the inflexion point of the isothermal magnetization curves and $T_{M0}$ was in previous studies [34,38] the value extrapolated from the transition field $B_{trans}$ at zero field. For the sake of simplicity, we will define here $T_{M0}$ as the FM-AFM transition temperature for low applied magnetic fields ($B$ =0.03-0.05 T). The saturation magnetization $M_S$ at different pressures $P$, was determined from the isothermal magnetization curves.

## III. RESULTS AND DISCUSSION

$Y_{0.7}Er_{0.3}Fe_2D_{4.2}$ compound crystallizes in the same monoclinic structure as $YFe_2D_{4.2}$ compound, described elsewhere in $Pc$ space group [28,38]. It is worth to recall that the Er for Y substitution induces a significant lattice contraction of about 0.6% [34,38]. The deuterium insertion together with the lowering of crystal symmetry has been found to have a large influence on the magnetic properties of the Fe sublattice magnetism[29].

### A. Isofield magnetization curves

The temperature dependence of the D.C. magnetization curves for $Y_{0.7}Er_{0.3}Fe_2D_{4.2}$ compound recorded at 0.03T are plotted in Fig. 1 for several applied pressures. A detailed look at the measurement performed at 0 GPa indicates a first regime with progressive increase of the magnetization from 2 K up to about 55 K. This temperature precisely coincides with $T_{Er}$, the temperature at which the Er sublattice magnetization vanishes as measured by neutron powder diffraction (NPD) [38]. This confirms that the Er sublattice exhibits a strong temperature reduction in $Y_{0.7}Er_{0.3}Fe_2D_{4.2}$. Another remarkable temperature is the inflection point derived for the ambient pressure measurement, a value easy to determine of $T_{inflexion} = 65(1)$ K, which precisely corresponds to $T_{M0} = 66$ K previously reported as the temperature at which the FM-AFM transition occurs [38]. We consequently can take these two remarkable points to investigate the change of magnetic state for the $Y_{0.7}Er_{0.3}Fe_2D_{4.2}$ compound. The values of $T_{Er}$ and $T_{M0}$ derived from the isofield magnetization curves recorded at 0.03T are listed in Table I. Looking at $T_{Er}$, the maximum of the bump in the magnetization curves plotted in Fig. 1, one can notice that the application of external pressure leads to a pronounced reduction of this critical temperature. Additionally, the height of the bump is also reduced for the larger pressures and this bump has already disappeared at pressure of 0.49 GPa. This means that applying external pressure permits to destroy the ferrimagnetic state in $Y_{0.7}Er_{0.3}Fe_2D_{4.2}$ compound. A similar behaviour occurs for $T_{M0}$ since the inflexion point has also disappeared for the largest studied pressure.

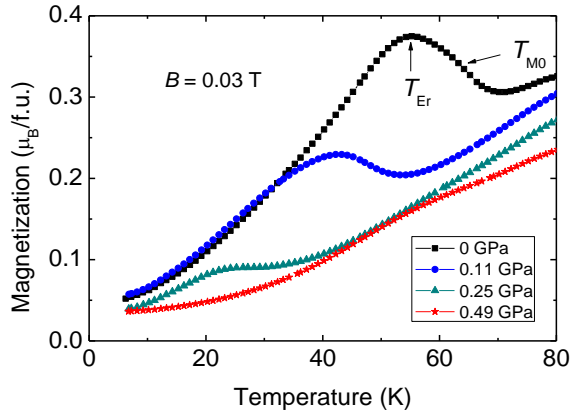

Fig 1: Comparison of the thermomagnetic behaviour of isofield magnetization curves recorded at 0.03 T for $Y_{0.7}Er_{0.3}Fe_2D_{4.2}$ compound at the indicated pressures ranging from 0 to 0.49 GPa.

The pressure dependence of both $T_{Er}$ and $T_{M0}$ has been followed by magnetic measurements (Fig. 2), they both rapidly decrease upon applying external pressure. A linear decrease of $T_{Er}$ and $T_{M0}$ versus $P$ (GPa) is obtained according to equations:

$$T_{Er}(K) = 56 - 129\,P \qquad (1)$$
$$T_{M0}(K) = 64 - 140\,P \qquad (2)$$

Using these equations one gets two critical pressures of $P_{crit}(Er) = 0.43 \pm 0.03$ GPa, pressure for which $T_{Er} = 0$ K and $P_{crit}(M0) = 0.46 \pm 0.02$ GPa, for $T_{M0} = 0$ K. This confirms the interpretation of the disappearance of both the Er magnetic ordering and the ferromagnetic ordering region of the Fe magnetic moments at pressures close to 0.49 GPa. Due to the uncertainty of this determination, one can expect the presence of a bicritical point at the merging point of the two lines, that is around 0.44 GPa ±0.04 GPa.

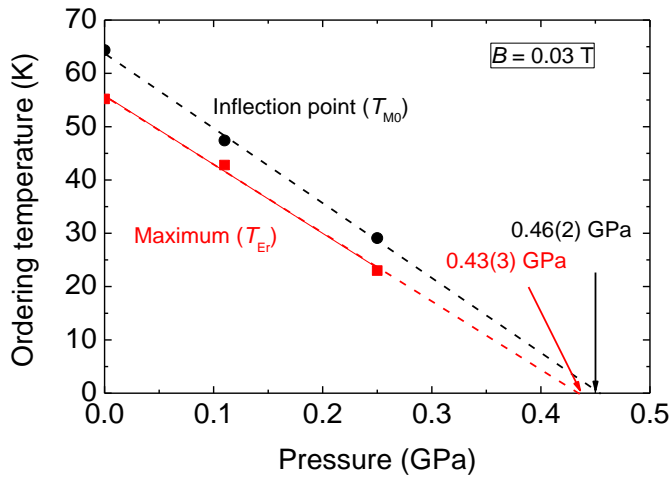

Figure 2 : Pressure induced decrease of the magnetic transition temperatures observed at 0.03 T for $Y_{0.7}Er_{0.3}Fe_2D_{4.2}$ in the pressure range 0 to 0.49 GPa.

Table I : Critical temperatures derived from isofield magnetization curves of $Y_{0.7}Er_{0.3}Fe_2D_{4.2}$ compound recorded at 0.03 T and 5 T in temperature range between 4 and 300 K for various pressures.

| $P$ (GPa) | $T_{Er}$ (K) @ 0.03T | $T_{M0}$ (K) @ 0.03T | $T_{Er}$ (K) @5T | $T_{M0}$ (K) @5T |
|---|---|---|---|---|
| 0 | 55 | 65 | 60 | 86 |
| 0.11 | 44 | 47 | 45 | 66 |
| 0.25 | 25 | 29 | 26 | 47 |

It is worth to compare the behaviour of $Y_{0.7}Er_{0.3}Fe_2D_{4.2}$ to that of $YFe_2D_{4.2}$ and $YFe_2H_{4.2}$ parent compounds [33]. For $Y_{0.7}Er_{0.3}Fe_2D_{4.2}$ compound the critical pressure $P_{crit}(M0)$ is significantly smaller than that of $YFe_2D_{4.2}$ (0.54 GPa) itself smaller than the pressure of 1.25 GPa estimated for $YFe_2H_{4.2}$. This most probably reflects the influence of the progressive unit cell expansion on the IEM of the Fe sublattice when going from $Y_{0.7}Er_{0.3}Fe_2D_{4.2}$ compound to $YFe_2D_{4.2}$ and then $YFe_2H_{4.2}$, since the cell volume has been demonstrated to play a key role on the Fe magnetism in these compounds [33]. The pressure sensitivity of $T_{M0}$ is of similar magnitude but slightly smaller for $Y_{0.7}Er_{0.3}Fe_2D_{4.2}$ compound ($dT_{M0}/dP = 140$ K GPa$^{-1}$) compared to that earlier reported for $YFe_2D_{4.2}$ ($dT_{M0}/dP = 156$ K GPa$^{-1}$) [35].

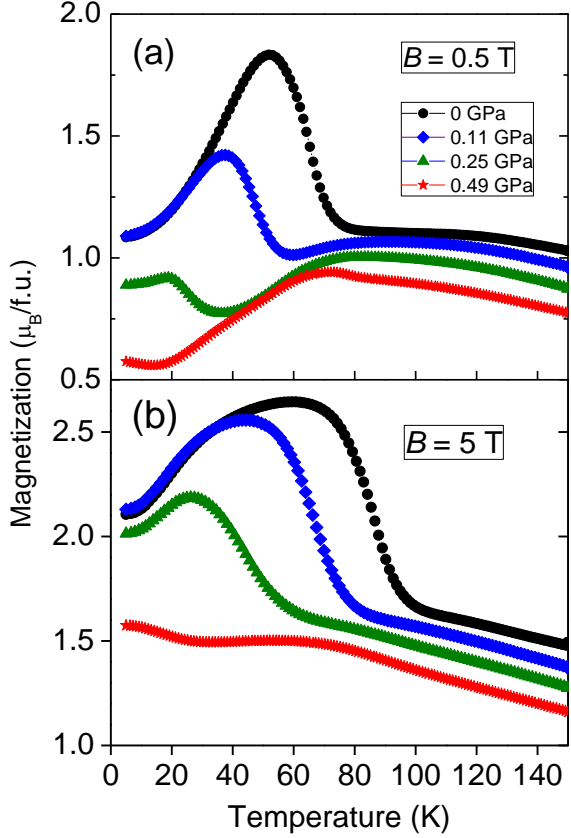

Figure 3 : Comparison of the thermomagnetic behaviour of isofield magnetization curves recorded at 0.5 T (a) and 5 T (b) for $Y_{0.7}Er_{0.3}Fe_2D_{4.2}$ compound at the indicated pressures ranging from 0 to 0.49 GPa.

Similar thermomagnetic curves under pressure have been recorded at different applied fields of 0.5 and 5 T and the results are plotted in Figures 3a and 3b respectively. The curves exhibit similar behaviour as the ones recorded at 0.03T. They are featured at low pressure by first an increase of the magnetization corresponding to the Er magnetic moment reduction up to a maximum magnetization value at $T_{Er}$, then followed by a strong decrease of the magnetization when approaching the $T_{M0}$ value corresponding to the reduction of the magnetization as expected for a FM-AFM transition. It is interesting that the 0.5 T curves provide transition temperatures (both $T_{Er}$ and $T_{M0}$) in close agreement with those determined from the 0.03 T isofield curves. They can consequently be considered as low magnetic field values. When applying a magnetic field an order of magnitude larger, that is 5 T, we can observe (Fig. 3b) that the temperature $T_{Er}$ corresponding to the maximum of the thermomagnetic curves is significantly shifted towards higher temperatures. In addition, it can be noticed that this maximum is much broader at 5 T than at low applied fields. The shift to higher values of $T_{Er}$ bears witness to the fact that the Er magnetic moment remains antiferromagnetically coupled to the Fe sublattice at larger temperature when increasing the magnetic field. It is also easily observable that $T_{M0}$ temperature is also raised as the applied magnetic field is increased to 5 T. This results from the reinforcement of the ferromagnetic coupling and leads to a wider temperature domain for the ferromagnetic ordering of the Fe magnetic moments. A value of $T_{M0}$ = 86 K is obtained at 5 T against 66 K for 0.03 T or 0.5 T. $T_{Er}$ is less affected by the application of an external magnetic field: $T_{Er}$ is larger of 5 K (10%) at ambient pressure and reduced of only 2 K at 0.25 GPa as the field increase from 0.03 to 5 T. Because of both $T_{M0}$ and $T_{Er}$ different pressure dependence, the FM temperature region is widened upon applying magnetic field on $Y_{0.7}Er_{0.3}Fe_2D_{4.2}$.

The pressure dependence of $T_{Er}$ and $T_{M0}$ transition temperatures at 5 T, are plotted in Fig. 4. They exhibit a linear decrease which can be fitted by the following relations:

$$T_{Er}(K) = 60 - 134\ P \qquad (3)$$
$$T_{M0}(K) = 86 - 156\ P \qquad (4)$$

A comparison with the equations 1 and 2 reveals that the slope is increased for both $T_{Er}$ (129 K.GPa$^{-1}$) and $T_{M0}$ (140 K.GPa$^{-1}$) of respectively 16 and 5 K.GPa$^{-1}$, meaning that the sensitivity of $T_{M0}$ to the external pressure is more reinforced by the larger applied magnetic field than $T_{Er}$.

By extrapolation of the curves plotted in Fig. 4, we can determine the critical pressure at which $T_{Er}$ and $T_{M0}$ are vanishing to 0 K: 0.45±0.01 GPa and 0.55 ±0.01 GPa respectively. The former critical pressure is much increased upon applying large magnetic field of 5 T in comparison with the 0.45 GPa value derived from the extrapolation of the low magnetic field curves plotted in Figs. 1 and 3a. This reveals that $T_{M0}$ is more sensitive to cell volume variation due to the IEM character of this transition, compared to $T_{Er}$. The slopes $dT/dP$ = -134 K and -156 K. GPa$^{-1}$ for $T_{Er}$ and $T_{M0}$ respectively are larger compared to those measured at 0.03 T. It is also remarkable that both the vanishing pressure and the slope measured at 5 T for $T_{M0}$ becomes similar to those measured at 0.03 T for $YFe_2D_{4.2}$ ($P_{Crit.}$ = 0.54 GPa and $dT/dP$ = -156 K GPa$^{-1}$). This is not really a coincidence: the AFM-FM transition temperature for $Y_{0.7}Er_{0.3}Fe_2D_{4.2}$ at 5 T is close to that of $YFe_2D_{4.2}$ extrapolated at 0 T ($T_{M0}$ = 84 K). This reveals a strong correlation between transition fields and the cell volume changes induced either by chemical substitution or by applied pressure at this IEM transition. The pressure induced decrease of $T_{M0}$ reflects that the AFM coupling within the Fe sublattice is favoured upon applied pressure because of the unit cell reduction. At high pressure ($P$ = 0.49 GPa) the absence of a maximum in the magnetization curves plotted for $Y_{0.7}Er_{0.3}Fe_2D_{4.2}$ (Fig. 3b) shows that no Er moment is subtracted to the Fe magnetization. This may indicate that either the Er magnetic moment is no longer coupled to the Fe one or else that it is also coupled in a way preserving the overall antiferromagnetic structure. Such assumption could be checked by neutron diffraction measurements under applied pressure.

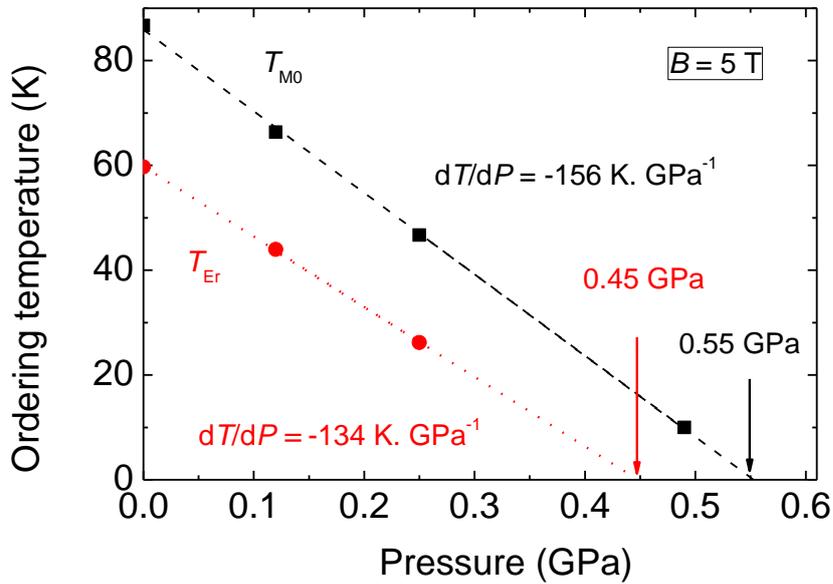

Fig. 4 : Pressure induced decrease of the magnetic transition temperatures observed at 5 T for $Y_{0.7}Er_{0.3}Fe_2D_{4.2}$ in the pressure range 0 to 0.49 GPa.

**B Isothermal magnetization curves**

Isothermal magnetization curves have been systematically recorded for $Y_{0.7}Er_{0.3}Fe_2D_{4.2}$ up to 7 T for different applied pressures ranging from 0 to 0.5 GPa. Examples of such magnetization curves are plotted in Fig. 5 for pressures of 0, 0.11 and 0.25 GPa. At low temperature such as 2 K, the magnetization curve is typical of ferromagnetic behaviour with a sharp increase at low field and tendency to saturation. However, the significant high field slope (above 2 T) is reminiscent of the ferrimagnetic nature of $Y_{0.7}Er_{0.3}Fe_2D_{4.2}$ compound, confirming its ground state. No remanent magnetization as well as no sign of significant coercivity field are observed here. As shown in Fig. 5, few change is observed upon heating from 2 to 10 K. Getting closer to $T_{Er}$ for instance at 40 and 50 K, the $M$(H) approach to saturation becomes faster than at lower temperature showing that the magnetization of $Y_{0.7}Er_{0.3}Fe_2D_{4.2}$ is easier to saturate because of the lower Er sublattice magnetization and of its vanishing magnetocrystalline anisotropy. Similar magnetization curves are observed in the ferromagnetically ordered region between $T_{M0}$ and $T_{Er}$ transition temperatures for instance at 60 K.

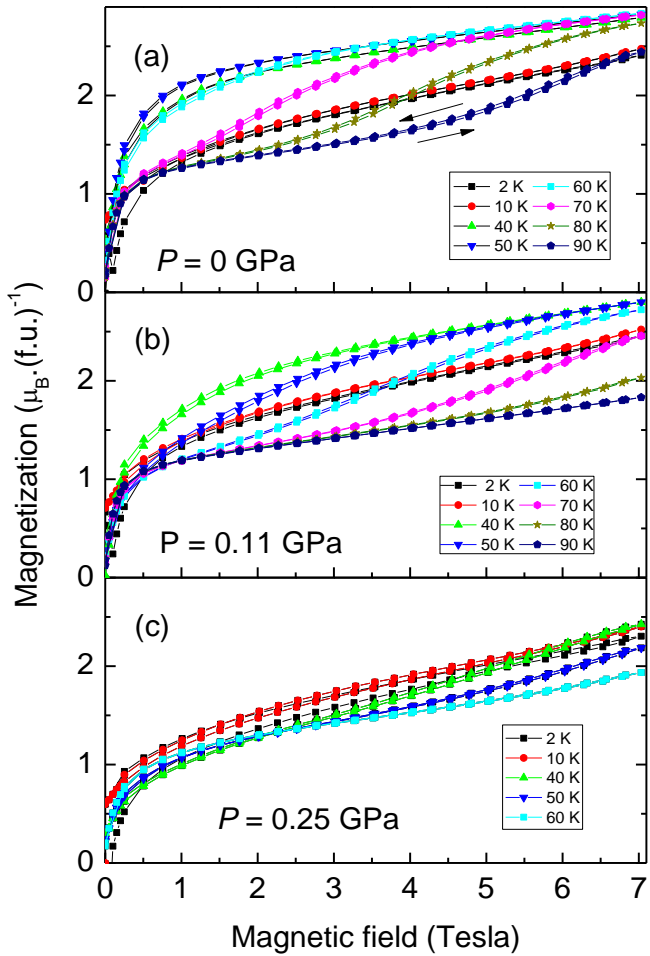

Fig. 5 : Isothermal magnetization curves recorded at (a) 0 GPa, (b) 0.15 GPa and (c) 0.25 GPa for $Y_{0.7}Er_{0.3}Fe_2D_{4.2}$ at the indicated temperatures between 2 and 90 K.

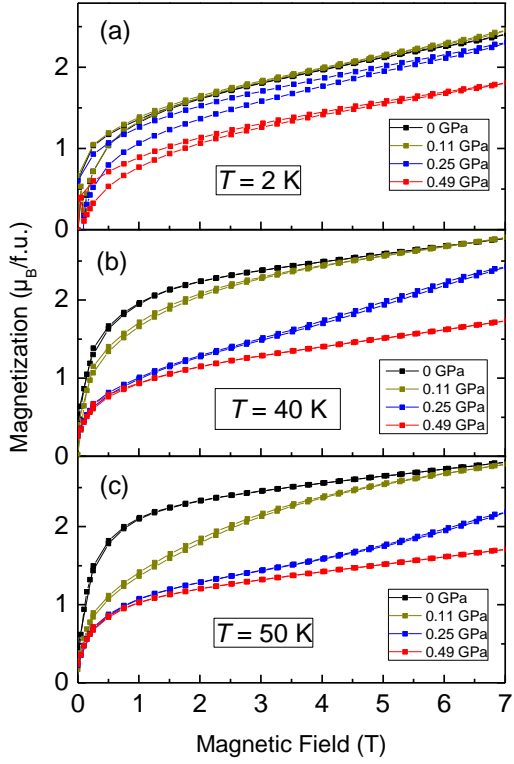

Fig. 6 : Isothermal magnetization curves recorded at 2 K (a), 40 K (b) and 50 K (c) for $Y_{0.7}Er_{0.3}Fe_2D_{4.2}$ at the indicated pressures.

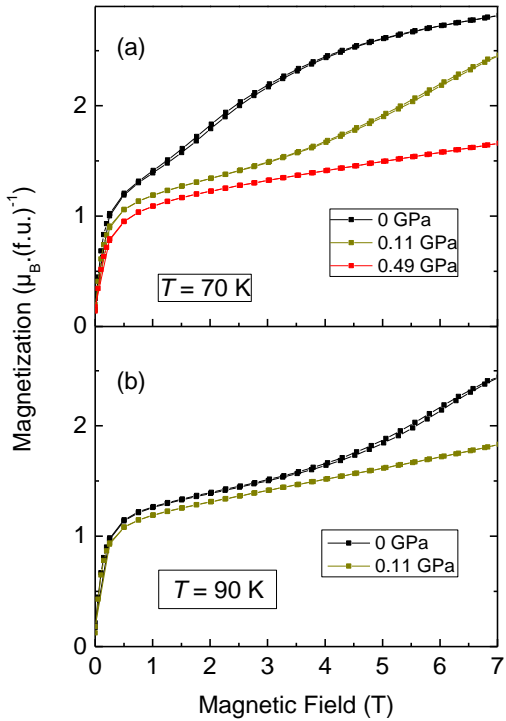

Fig. 7: Isothermal magnetization curves recorded at 70 K (a) and 90 K (b) for $Y_{0.7}Er_{0.3}Fe_2D_{4.2}$ at the indicated pressures.

Above $T_{M0}$ the AFM ordering of the compound leads to very different magnetic behaviours. Indeed, the $M(B)$ curve proceeds in two main parts: first a sharp increase up to about 1.25μ$_B$ f.u$^{-1}$ where a pronounced slopy plateau occurs and a second increase at larger field. This last increase is a fingerprint of the itinerant electron metamagnetic behaviour reported earlier in YFe$_2$D$_{4.2}$ and YFe$_2$H$_{4.2}$ [29] as well as Y$_{0.7}$Er$_{0.3}$Fe$_2$D$_{4.2}$ compound [38]. This is typical of the 3d electron magnetism of the Fe sublattice and reflects the AFM to FM transition as induced by the external applied magnetic field. It is noteworthy that the transition field of this IEM transition $B_{trans}$ is fast shifted to larger values upon increasing temperature. This is clearly seen from Fig. 5b where the transition field $B_{trans}$, determined as the inflection point of the isothermal magnetization curve is increasing from 70 to 80 and then 90 K. The influence of the temperature and magnetic field on $B_{trans}$ of Y$_{0.7}$Er$_{0.3}$Fe$_2$D$_{4.2}$ has been detailed in [38] based on high magnetic field measurements. It is interesting to remark that Y$_{0.7}$Er$_{0.3}$Fe$_2$D$_{4.2}$ does not exhibit large hysteresis cycle in contrast to YFe$_2$D$_{4.2}$ and YFe$_2$H$_{4.2}$ which present significant hysteresis cycles at low temperatures [33]. The reason for the almost disappearance of the hysteresis at the IEM transition in Y$_{0.7}$Er$_{0.3}$Fe$_2$D$_{4.2}$ may originate from the reduced unit cell volume as induced by the Er for Y substitution since the IEM is known to be very sensitive to the cell volume [33].

The isothermal magnetization curves recorded at $P$ = 0.11 GPa are gathered for different temperatures in Fig. 5b. They present a similar behaviour as that observed at lower pressure. The noticeable changes are smaller magnetization values at low field, indicating a more difficult magnetization process. At $P$= 0.11 GPa, the 40 and 50 K magnetization curves differ significantly. The first one reflects the ferrimagnetic ground state whereas at 50 K an inflection point is seen able confirming that $T_{FM-AFM}$ < 50 K < $T_N$. More pronounced inflections of the curves are observed at 60 and 70 K, confirming the existence of this IEM transition. For $T$= 80 and 90 K, the transition field $B_{trans}$ is shifted further away to applied field larger than the 7 T used here.

The magnetization curves recorded at a pressure of 0.25 GPa are plotted in Fig. 5c. At a first glance, they look very similar however two different behaviours can be distinguished. The ferrimagnetic type one recorded at 2 and 10 K shows a continuous increase of magnetization and an even larger high field susceptibility than for lower pressures. At 40K and above the magnetization curves exhibits a transition field at the IEM transition. This transition just starts at 60 K since the transition fields are shifted to field higher than 7 T.

Figs. 6 to 7 give a comparison of the magnetization curves recorded at different pressures but at identical temperatures between 2 and 90 K. At 2 K, the main effect of the pressure is a decrease of the saturation magnetization and a hysteresis between the curves recorded upon increasing and decreasing magnetic field (Fig. 8). This hysteresis increases from 0 to 0.25 GPa, that is in the region where Er is magnetically ordered, but is reduced as observed in the FM state at 0.49 GPa, when only the Fe sublattice plays a role. We conclude that the main part of the observed hysteresis arises from the Er sublattice, being most probably of magnetocrystalline anisotropy origin. This interpretation is further confirmed by the noteworthy absence of hysteresis in the 40 and 50 K magnetization curves. At 60 K and above the $M$(B) curves recorded upon increasing and decreasing the magnetic field are also found to be barely identical. The absence or very small hysteresis observed here for $Y_{0.7}Er_{0.3}Fe_2D_{4.2}$ contrasts with the large hysteresis of more than 1 T wide reported at low temperature for the parent $YFe_2D_{4.2}$ or $YFe_2H_{4.2}$ compounds [33]. This surprising phenomenon is most probably related to reduction of the unit cell occurring upon Er for Y substitution. Indeed, similar reduction of the hysteresis cycle observed for the IEM has been reported upon pressure induced reduction of the unit cell of $YFe_2D_{4.2}$ or $YFe_2H_{4.2}$ compounds. This demonstrates that the presence of Er has a significant effect on the IEM behaviour of the Fe sublattice via the unit cell volume reduction.

In order to study the influence of pressure on the magnetization, we have calculated the saturation magnetization $M_s$ and the coefficient $dln(M_s)/dln(P)$= -0.31(1) upon heating from the analysis of the magnetization curves recorded at 2 K for several pressures (Fig. 8). In addition a linear decrease of the average magnetization $M_a$ between heating and cooling $dln(M_a)/d(P)$= -1.06(4) GPa$^{-1}$ cooling was observed above 0.1 GPa.

A comparison of the magnetization curves recorded at different pressures at 40 K in Fig. 6b and 50 K in Fig. 6c shows that the classical magnetization behaviour observed at ambient pressure is modified upon application of pressure and transforms towards S shape magnetization process typical of IEM transition. At this point one can recall, that the measurement being performed on polycrystalline sample, the transition is expected to be smeared out in comparison to what could be expected from single crystal measurements. At 70K or 80 K even the ambient pressure, magnetization curves exhibit a metamagnetic transition (Fig. 7). The application of pressure leads to a large shift of the transition fields toward higher magnetic field values.

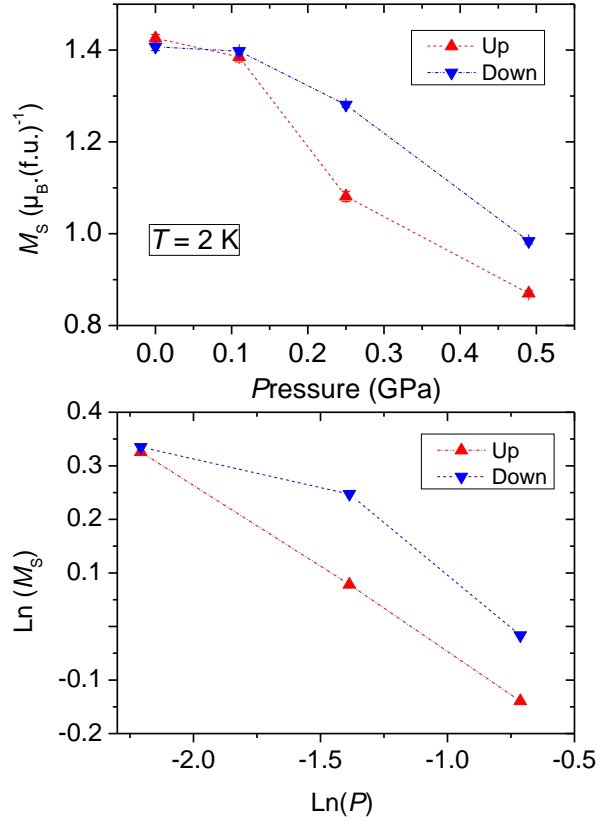

Fig. 8: Saturation magnetization at 2 K versus applied pressure upon increasing (up) and decreasing field (down). Below ln($M_S$) = f(ln$P$).

Taking the inflection point of the magnetization curves to determine the critical field we have plot the pressure dependence of $B_{trans}$ versus temperature in Fig. 9 at 0 and 0.11 GPa. This has been done at several different temperatures. $B_{trans}$ increases linearly versus temperature and is shifted to lower values upon an applied pressure of 0.11 GPa. The slope is similar upon 0.11 GPa than without pressure (0.22 T.K$^{-1}$). The extrapolation to zero field leads to $T_{M0}$ = 43 K for 0.11 GPa and 60 K for 0 GPa, which confirms the high sensitivity of the FM-AFM transition temperature related to the pressure induced cell volume change.

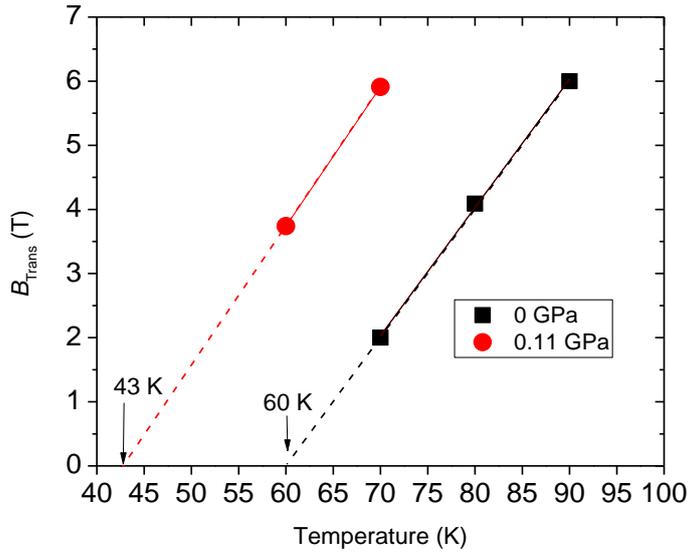

Fig. 9: Temperature dependence of the critical field $B_{trans}$ of the IEM transition as derived from the inflection point of the isothermal magnetization curves of $Y_{0.7}Er_{0.3}Fe_2D_{4.2}$ recorded at the indicated pressures.

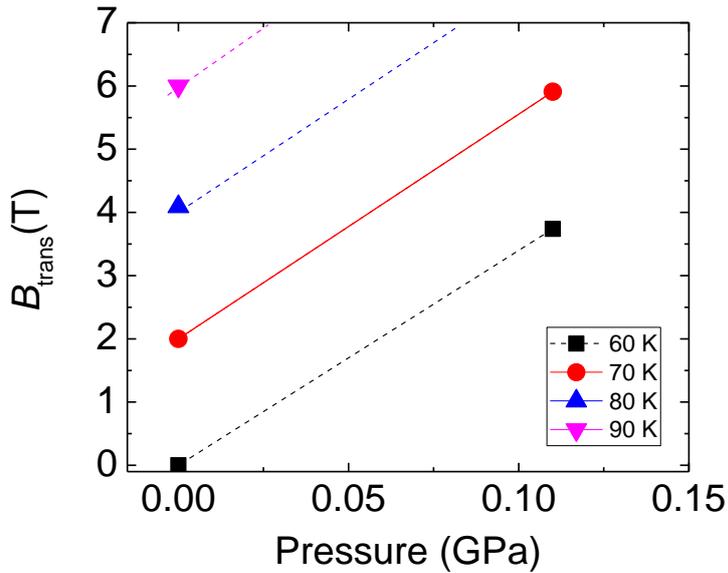

Fig. 10: Evolution of the critical field $B_{trans}$ of the IEM transition versus the applied pressure, as deduced from isothermal magnetization curves of $Y_{0.7}Er_{0.3}Fe_2D_{4.2}$ recorded at the indicated temperatures. Lines are guide for the eyes.

Fig. 10 presents the evolution of the critical field $B_{trans}$ of the IEM transition versus the applied pressure for $Y_{0.7}Er_{0.3}Fe_2D_{4.2}$. It shows that the transition field increases versus temperature whatever the applied pressure, at 80 and 90 K for 0.11 GPa becoming larger than 7 T by extrapolation. $B_{Trans}$ also increases versus applied pressure for a given temperature, in line with the lowering of the FM-AFM transition temperature.

## C. Discussion

The magnetic properties of YFe$_2$ can be tuned under hydrostatic pressure as a collapse of the Fe moment has been observed above 90 GPa by nuclear forward scattering (NFS) of synchrotron radiation as reported by Lübbers et al. [41,42]. This collapse was attributed to the existence of a Fe-Fe critical distance of 2.30 Å, below which the Fe becomes nonmagnetic. Later, Zhang et al.[43] explained the collapse of the Fe moment in $R$Fe$_2$ compounds ($R$ = Y, Lu, Hf and Zr) by *ab-initio* Density functional theory (DFT) calculations as the 3d band broadening under pressure reduces the splitting of the majority and minority bands and therefore the Fe moment. At high pressure a quantum transition to a low or zero spin state was observed. A small decrease of the YFe$_2$ moment per unit mass dln$M$/d$P$ = -8.4(4) 10$^{-5}$ GPa$^{-1}$ was obtained using forced magnetostriction measurements by Armitage et al.[44]. According to Lubbers et al. [41] the Curie temperature of C15 YFe$_2$ first increases from 535 K to 660 K under 15 GPa and decreases for higher pressure. Above 20 GPa YFe$_2$ adopts a hexagonal C14 structure, and an FM-AFM transition is observed around 50 GPa, when the Fe 2$a$ sites becomes nonmagnetic like in C14 ScFe$_2$. As previously mentioned, there is no more ordered Fe moment above 90 GPa.

It is interesting to compare this evolution of the magnetic properties of YFe$_2$ under hydrostatic pressure with that of YFe$_2$D$_x$ deuterides versus D content at ambient pressure as they present similar features: an increase of $T_C$ up to 720 K for $x$ =1.2, followed by a decrease of $T_C$ to 300 K for $x$ = 3.5, a FM-AFM transition for $x$ = 4.2 and a collapse of the Fe moment for $x$ = 5 [15]. This is very surprising if only Fe-Fe distance contraction is considered as a driving force for the magnetic properties of YFe$_2$, as D (H) insertion increases continuously the cell volume up to 26 % for $x$ = 5 and therefore the Fe-Fe distances. Indeed, the contribution of the s electrons originating from H or D atoms which modify the DOS and strengthen the itinerant character of the Fe moments should be also taken into account [30]. A dominating volume effect on the Fe moment is observed up to $x$ =3.5 as it yields an increase of the Fe moment due to a better localization of the 3d Fe band, but for larger H(D) content the influence of Fe-H bonding becomes predominant, and a sharp decrease of the Fe moment is observed both experimentally and by *ab-initio* calculations [30]. Therefore, we can conclude that H or D insertion cannot be considered as a simple negative pressure effect and that the modification of the electronic properties, in particular the DOS at E$_F$ are of utmost importance to understand the magnetic properties of the YFe$_2$ hydrides and deuterides.

Then, to explain the specific magnetic behaviour of YFe$_2$(H,D)$_{4.2}$ compounds one should consider not only the cell volume change and the Fe-H bonds, but also the lowering of crystal symmetry from cubic to monoclinic structure due to the ordering of H(D) atoms into tetrahedral interstitial sites [26,30,45]. In the monoclinic structure, the Fe sites are no more equivalent as there is 8 different Fe sites instead of 1 Fe site in the cubic C15 phase [28]. This yields a broad distribution of Fe-Fe distances and of local Fe moments as observed by $^{57}$Fe Mössbauer spectroscopy [32]. According to NPD experiments, the FM structure is constituted of parallel Fe moments oriented perpendicular to the monoclinic *b* axis [29]. The AFM structure can be described by the stacking of antiparallel ferromagnetic Fe layers, with Fe moments also perpendicular to the monoclinic *b* axis [29]. These antiparallel ferromagnetic layers are separated by an intermediate non-magnetic layer and the AFM magnetic cell is described with a doubling of the *b* cell parameter compared to the FM cell. The FM-AFM transition has been explained by the IEM behavior of the Fe sublattice, monitored by the loss of ordered moment on a particular Fe site inducing an inversion of its Fe neighbour's moment orientation. The crystallographic origin of this FM-AFM transition is comparable to that of C14 hexagonal *R*Fe$_2$ compound (ScFe$_2$ under pressure, (Hf,Ta)Fe$_2$) [41,46,47] where the collapse of the Fe moment on the 2*a* site but not on the 6h site stabilizes the AFM structure at the expense of the FM one. This first order transition is also accompanied by a cell volume contraction at the transition and a linear increase of $T_{M0}$ versus applied field, which is currently observed in first order IEM systems [48].

As already indicated, the FM-AFM transition temperature $T_{M0}$ is very sensitive to cell volume changes as expected from an IEM behavior. The cell volume reduction related to the hydrostatic pressure also yields a decrease of $T_{M0}$ for both YFe$_2$D$_{4.2}$ and YFe$_2$H$_{4.2}$ and reveals the existence of a common critical cell volume ($V_0$ = 501.7(3) Å$^3$) for the onset of ferromagnetism [33]. The Er for Y substitution reduces chemically the cell volume of the hydrides and deuterides and induces a linear decrease of $T_{M0}$[34]. However, it was observed that the reduction of $T_{M0}$ cannot be explained by a pure volume effect as different variations of $T_{M0}$ versus cell volume were observed: d$T_{M0}$/d$V$ = 16.4 K Å$^{-3}$ and 15.9 K Å$^{-3}$ for YFe$_2$D$_{4.2}$ and YFe$_2$H$_{4.2}$ respectively under applied pressure, d$T_{M0}$/d$V$ = 13.1(2) K Å$^{-3}$ upon D for H substitution and d$T_{M0}$/d$V$ = 6.3(2) K Å$^{-3}$ and 6.7(2) K Å$^{-3}$ upon Er for Y substitution for hydrides and deuterides respectively (see Fig. 10(b) in [34]). The first aim of this study was therefore to observe the combined influence of Er substitution and applied pressure on the FM-AFM transition. For this purpose, Y$_{0.7}$Er$_{0.3}$Fe$_2$D$_{4.2}$ deuteride, which was well characterized at ambient pressure by NPD and magnetic measurements, has been selected [38]. Considering d$T_{M0}$/d$P$ = 140 K GPa$^{-1}$ measured at 0.03 T, assuming the same compressibility $\kappa = 0.013$ GPa$^{-1}$ than for YFe$_2$D$_{4.2}$ [35] and the cell volume of Y$_{0.7}$Er$_{0.3}$Fe$_2$D$_{4.2}$ at 300 K $V_0 = 500$ Å$^3$ yields d$T_{M0}$/d$V = -21.5$ K Å$^{-3}$. This variation is larger than for YFe$_2$D$_{4.2}$ and YFe$_2$H$_{4.2}$ meaning that there is a synergetic effect between chemical and hydrostatic pressure effects. The extrapolated critical cell volume $V = 497$ Å$^3$ below which the FM structure becomes less stable than the AFM one is smaller than that calculated for YFe$_2$D$_{4.2}$ and YFe$_2$H$_{4.2}$ under pressure (501.7(3) Å$^3$). Note that, the Er substitution at ambient pressure yield to a critical volume of 490.4 Å$^3$ for the onset of ferromagnetism in the deuterides, i.e; much smaller than under pressure [34]. This means that even if there is no more ordered Er moment at $T_{M0}$, local Er-Fe interactions are still effective to stabilize the FM state at a lower cell volume.

A second objective of this study was to observe how the hydrostatic pressure influences the Er-Fe interactions. Y$_{0.7}$Er$_{0.3}$Fe$_2$D$_{4.2}$ is ferrimagnetic at low temperature but undergoes a forced ferrimagnetic-ferromagnetic transition at an applied field of 8 T, which compared to other R-Fe systems [21,49], is remarkably moderate and indicates that D absorption weakens significantly the Er-Fe interactions due to the large increase of Er-Fe distances [38]. In this work, the influence of pressure on Er-Fe interaction was followed through the variation of the Er magnetic ordering temperature $T_{Er}$ under applied pressure. Experimentally, at 0.03 T the variation of $T_{Er}$ and $T_{M0}$ versus pressure are quite close and converge towards the same critical pressure, however the difference of d$T$/d$P$ becomes larger under an applied field of 5 T. This can be explained by the linear increase of $T_{M0}$ versus applied field, whereas $T_{Er}$ is not very sensitive to the applied field[38]. Indeed, it has been observed that the transition field from forced ferri to ferromagnetic state remains almost constant versus temperature up to $T_{Er}$, and also versus Er content in Y$_{1-x}$Er$_x$Fe$_2$D$_{4.2}$ deuterides ($B_C = 8$ T) [34]. This reveals a weak sensitivity of the mean Er-Fe interaction $J_{Er-Fe}$ to the applied field and Er content. However, in the present study the Er sublattice becomes paramagnetic under pressure because of the decoupling of the Er and Fe sublatices.

The saturation magnetization $M_S$ of Y$_{0.7}$Er$_{0.3}$Fe$_2$D$_{4.2}$ at 2 K decreases under pressure as dln($M_S$)/d$P$ =-1.06(4) GPa$^{-1}$ (Average value of $M_s$ in µ$_B$ f.u$^{-1}$), this value is significantly larger than observed for YFe$_2$D$_{4.2}$ (dln($M_S$)/d$P$ =-7.3 10$^{-2}$ GPa$^{-1}$) [35] and can be attributed to Er ordered moment reduction. At 5 T the deuteride is still in a ferrimagnetic state, and the $M(B)$ curves displays a large slope characteristic of the Er anisotropy. At ambient pressure, the average Er moment refined from the NPD pattern was of only 6.5 µ$_B$ compared to 9 µ$_B$ for the free ion value, this reduction was previously discussed and attributed to a crystal field effect[38]. The reduction of the Er contribution can be therefore attributed either to a larger crystal field effect or to a disorder of the Er moment orientation under pressure. Neutron diffraction under pressure will be necessary to follow independently the evolution of both Er and Fe moments, but it is beyond the scope of this study. In $ErFe_2$ at 4.2 K, $\mu_{Er}$ = 8.47 $\mu_B$ and $\mu_{Fe}$= 1.97 $\mu_B$ as determined by NPD and both sublattice order ferrimagnetically below $T_C$ = 600 K with a compensation temperature at $T_{comp}$ = 400 K [50]. The Er for Y substitution in induces a decrease of $T_{comp}$ which reach 240 K for $Y_{0.5}Er_{0.5}Fe_2$. This clearly reveals that a decrease of the Er-Fe interactions and a reduction of the Er moment due to crystal field effect was considered [51]. Deuterium absorption in $ErFe_2$ also decouple the ordering temperatures of both sublattices as observed in previous NPD study of $ErFe_2D_{3.5}$ ($\Delta V/V$ = 14.5 %) [52,53]. It showed that the Er sublattice is more affected by D absorption than the Fe sublattice with a decrease of both $T_{Er}$ = 300 K compared to $T_{Fe}$ =450 K and a reduction of $\mu_{Er}$ = 4.3 $\mu_B$ compared to 9 $\mu_B$ whereas $\mu_{Fe}$ remains constant [52,53]. Further magnetic studies of $ErFe_2H_x$ hydrides indicated they are ferrimagnetic with a decrease of $T_{comp}$ versus H content, corresponding to a reduction of the molecular field and therefore the Er-Fe interactions [54]. All these studies show that the Er ordering temperature is more sensitive to the dilution by Y or the insertion of hydrogen than the Fe sublattice, and that crystal field effects occur for large H(D) content. In $Y_{0.7}Er_{0.3}Fe_2D_{4.2}$, both Y dilution and large D content can explain the strong weakening of the Er-Fe interactions which occurs through the hybridization of the 5d and 3d orbitals [55].

It could be interesting to compare the results obtained in this work with the influence of hydrostatic pressure on the magnetic properties of $ErCo_2$ [56] and $Y_{1-y}Er_yCo_2$ [57] Laves phase compounds as the Co sublattice also presents an IEM behavior. These Laves phases crystallize in a cubic C15 structure, but a rhombohedral distortion ($R\bar{3}m$) S.G.) with one Er site and 2 Co sites has been observed by NPD below $T_C$ for $ErCo_2$ [56]. The pressure induces an anisotropic cell reduction with $\kappa_a$ = 0.0051 GPa$^{-1}$ and $\kappa_c$ = 0.0076 GPa$^{-1}$ at 10 K for $ErCo_2$. In $ErCo_2$ a decoupling of the $T_{Co}$ and $T_{Er}$ ordering temperatures is observed under pressure. $T_{Er}$ remains almost constant with $dT_C(Er)/dP$ > 0.3 K GPa$^{-1}$ whereas $T_{Co}$ decreases with a rate of -3.45(3) K GPa$^{-1}$. In addition, in $ErCo_2$ the low temperature Er moment magnitude remains constant under pressure whereas the Co moment decreases with $dM_{Co}/dP$ = -0.1 $\mu_B$ GPa$^{-1}$. However, when Er is partially replaced by Y, in $Y_{0.3}Er_{0.7}Co_2$ Laves phase both Er and Co transition temperatures are reduced under pressure, but with a larger decrease for $T_{Co}$ compared to $T_{Er}$ [57].

In $Y_{0.7}Er_{0.3}Fe_2D_{4.2}$ $T_{Er}$ is smaller than $T_{M0}$ and the decrease of $T_{Er}$ versus applied pressure at low field remains close to that of $T_{M0}$. This can be related to the different nature of 3d metal magnetism between $RFe_2$ and $RCo_2$ Laves phases. For instance, $YCo_2$ and $LuCo_2$ are known as exchanged enhanced paramagnetic materials [58] whereas the corresponding YFe$_2$ compound has a ferromagnetic Fe order [6]. The Fe containing phases are featured by intrinsic ordered Fe magnetic moments unlike the $R$Co$_2$ whose Co sublattice presents magnetic induced fields by the exchange with magnetic rare earth[59]. Therefore, although both Y$_{0.7}$Er$_{0.3}$Fe$_2$D$_{4.2}$ and Y$_{1-y}$Er$_y$Co$_2$ compounds display an IEM behaviour, they have not the same origin and their Er and transition metal sublattice do not display the same sensitivity to applied pressure.

## IV. CONCLUSIONS

Hydrogen and deuterium insertion in YFe$_2$ compound leads to a cell volume expansion and a modification of the DOS which surprisingly present several similarities with the influence of applied pressure on the parent compound concerning the evolution of their magnetic properties (variation of $T_C$, FM-AFM transition, collapse of the Fe moment). For a critical H(D) content of 4.2 H(D)/f.u. a first-order FM-AFM transition very sensitive to cell volume variation has been observed in YFe$_2$H(D)$_y$. This transition presents many characteristics of an IEM behaviour, and its temperature can be tuned by applying a hydrostatic pressure, by H(D) isotope effect or by chemical substitutions on the Y site, as observed in different works in particular for Y$_{1-x}$Er$_x$Fe$_2$H(D)$_{4.2}$ compounds[34,38,39]. In this study we have, for the first time, combined both hydrostatic and chemical pressure by measuring the magnetic properties under pressure of Y$_{0.7}$Er$_{0.3}$Fe$_2$D$_{4.2}$ deuteride. This compound was selected as its crystal structure and magnetic properties were fully characterized by neutron diffraction and high magnetic field measurements and because the two ordering temperatures are well separated.

The shape of the $M(T)$ magnetization curves allows to identify two magnetic transition temperatures, the first one corresponding to the Er magnetic ordering temperature $T_{Er}$ (maximum of the magnetization, $T_{Er}$ = 55 K at ambient pressure) and the second one $T_{M0}$ due to the FM-AFM transition of the Fe sublattice (inflexion point, $T_{M0}$ = 66 K at ambient pressure). Both temperatures decrease linearly versus applied pressure with two different d$T$/d$P$ slopes and converge to a bicritical pressure of 0.44 ±0.04 GPa for a weak applied field of 0.03T above which the Er moments are no more ordered and the Fe sublattice adopts an antiferromagnetic structure. Interestingly, the difference between these two critical pressures increases as the applied field is raised up to 5 T with $P_{Crit.}$= 0.45 and 0.55 GPa for Er and Fe respectively. Indeed, the ordering of the Er moment is less sensitive to the applied pressure and applied field, than the FM-AFM transition temperature which varies in an opposite direction: $T_{M0}$ decreases with the applied pressure but increases with the applied field.

Compared to YFe$_2$D$_{4.2}$, $T_{M0}$ is smaller at $P = 0$ GPa and 0.03 T, and less pressure sensitive in Y$_{0.7}$Er$_{0.3}$Fe$_2$D$_{4.2}$ compound as observed from the differences of d$T_{M0}$/d$P$ slopes. But as the magnetic field increases to 5 T, the behaviour of Y$_{0.7}$Er$_{0.3}$Fe$_2$D$_{4.2}$ becomes close to that of YFe$_2$D$_{4.2}$ at 0.03T revealing a strong correlation between cell volume and applied field variation related to the IEM behaviour. The decrease of the saturation magnetization versus applied pressure at 2 K, can be mainly attributed to the Er magnetic sublattice, when compared to YFe$_2$D$_{4.2}$ under pressure. Above $T_{M0}$, the magnetization curves display a metamagnetic behaviour, which transition field $B_{trans}$ increases linearly versus temperature. The applied pressure shifts systematically $B_{trans}$ to lower temperature.

These results show that in Y$_{0.7}$Er$_{0.3}$Fe$_2$D$_{4.2}$, the Er and Fe magnetic sublattices present different ordering temperatures which are both very sensitive to the applied pressure, although they are decoupled and do not correspond to the same type of interactions. The comparison of these results with those of the literature concerning either YFe$_2$ or Y$_{1-y}$Er$_y$Co$_2$ compounds under pressure indicates that the magnetic properties of Y$_{0.7}$Er$_{0.3}$Fe$_2$D$_{4.2}$ depend not only on the cell volume changes but also on the influence of the Fe-D bonding and the lowering of crystal symmetry induced by long range deuterium order.

**Acknowledgement:** The financial support of Grant Agency of the Czech Republic (Grant No. 15 – 03777S) is acknowledged. The authors thanks also the CNRS, University of Grenoble and University of Paris East for financial support.